\documentclass[12pt]{article}
\usepackage{jheppub}
\usepackage{graphicx,color}
\usepackage[dvipsnames]{xcolor}
\usepackage{braket}
\usepackage{soul}
\usepackage{amssymb,amsmath}
\usepackage{mathtools}
\usepackage{bbm}
\usepackage{array}
\usepackage{simpler-wick}
\usepackage{verbatim}
\usepackage{anyfontsize}
\usepackage{dsfont}
\usepackage{tikz}
\usepackage{subcaption}
\usepackage{sidecap}
\usepackage{ytableau}
\sidecaptionvpos{figure}{t}
\usepackage[english]{babel}
\usepackage{enumitem}  
\usepackage[export]{adjustbox}
\usepackage{bbold}
\allowdisplaybreaks
\usepackage[]{hyperref}
\hypersetup{
    colorlinks = true,
    allcolors = {ForestGreen}
}
\allowdisplaybreaks

\newcommand{\nn}{\nonumber\\}
\DeclareMathOperator{\tr}{tr}

\def\d{\mathrm{d}}
\def\i{\mathrm{i}}
\def\e{\mathrm{e}}

\renewcommand{\Im}{{\rm Im}}
\renewcommand{\Re}{{\rm Re}}

\def\del{\partial}

\def\eqref#1{(\ref{#1})}

\def\be{\begin{equation}}
\def\ee{\end{equation}}

\numberwithin{equation}{section}
\let\mc=\mathcal
\let\mb=\mathbb

\newcommand{\inv}{\frac{1}}

\renewcommand{\O}{{\mathcal{O}}}

\newcommand{\de}{\delta}

\newcommand{\SL}{\mbox{SL}(2,\mathbb{R})}

\newcommand{\La}{\Lambda}

\renewcommand{\l}{\lambda}

\newcommand{\bpsi}{\bar{\psi}}

\title{On random matrix statistics of 3d gravity}

\author{Daniel L. Jafferis,}
\author{Liza Rozenberg,}
\author{Debmalya Sarkar,}
\author{Diandian Wang}

\affiliation{Jefferson Physical Laboratory, Harvard University, Cambridge, MA 02138, USA}
\emailAdd{jafferis@g.harvard.edu}
\emailAdd{erozenberg@g.harvard.edu}
\emailAdd{dsarkar@g.harvard.edu}
\emailAdd{diandianwang@fas.harvard.edu}

\date{\today}

\begin{document}

\abstract{
    We show that 3d gravity on manifolds that are topologically a Riemann surface times an interval  $\Sigma_{g,n}\times I$ with end-of-the-world branes at the ends of the interval is described by a random matrix model, namely the Virasoro minimal string. Because these manifolds have $n$ annular asymptotic boundaries, the path integrals naturally correspond to spectral correlators of open strings upon inverse Laplace transforms.
    For $g=0$ and $n=2$, we carry out an explicit path integration and find precise agreement with the universal random matrix expression. For Riemann surfaces with negative Euler characteristic, we evaluate the path integral as a gravitational inner product between states prepared by two copies of Virasoro TQFT. Along the way, we clarify the effects of gauging the mapping class group and the connection to chiral 3d gravity.
}

\maketitle

\section{Introduction}

The gravitational path integral is notoriously difficult to evaluate in general. In three-dimensional pure Einstein gravity with a negative cosmological constant (3d gravity), however, substantial progress has been made. For manifolds admitting a hyperbolic structure, the formulation in terms of Virasoro topological quantum field theory (Virasoro TQFT) \cite{Collier:2023fwi,Collier:2024mgv} provides a conceptually simple method for computing the path integral at finite gravitational coupling. Such manifolds have played an important role in revealing the ensemble nature of CFTs dual to 3d gravity \cite{Kraus:2016nwo,Cardy:2017qhl,Collier:2019weq,Belin:2020hea,Belin:2021ryy,Anous:2021caj,Kusuki:2021gpt,Schlenker:2022dyo,Numasawa:2022cni,Chandra:2022bqq,deBoer:2023vsm,deBoer:2024mqg,Wang:2025bcx,Hung:2025vgs,Jafferis:2025yxt,Chandra:2025fef,Hartman:2025ula,Liu:2025ikq}. 

Extrapolating Virasoro TQFT beyond hyperbolicity turns out to be not always well-defined \cite{Collier:2023fwi,Yan:2025usw}. Nevertheless, one can still perform the path integral directly, at least in principle. The path integral of 3d gravity on the torus wormhole $T^2\times I$, with asymptotic AdS boundaries at the ends of the interval, has been computed by Cotler and Jensen \cite{Cotler:2020ugk}, and the result displays random matrix behavior, providing further insight into the ensemble nature of 3d quantum gravity.

To establish an exact correspondence between 3d gravity and random matrices, however, requires knowledge of the gravitational path integral on manifolds with more complicated topology. In \cite{Jafferis:2024jkb}, it was conjectured that 3d gravity on manifolds that are topologically a Riemann surface times a circle, $\Sigma_{g,n}\times S^1$, is described by an infinite set of matrix models, one for each spin sector. Performing the path integral for general $g$ and $n$ would be a significant step in this direction, but it remains a challenge.

To this end, a simpler setup is to consider 3d gravity with end-of-the-world (EOW) branes. On-shell, 3d gravity with EOW branes computes OPE statistics of boundary CFTs (BCFTs) \cite{Wang:2025bcx,Hung:2025vgs}, but less is known about off-shell manifolds in this theory. In \cite{Jafferis:2025yxt}, it was conjectured that the gravitational path integral on $\Sigma_{g,n}\times I$, with EOW branes at the ends of the interval, is dual to a single matrix model, whose density of states is given by the Cardy density. This matrix model is itself also dual to the Virasoro minimal string (VMS) \cite{Collier:2023cyw}, so we will use the term ``VMS" to mean both the matrix model and the worldsheet theory.

In this paper, we prove this conjecture by computing the corresponding gravitational path integrals. We will first review how the disk amplitude is reproduced in 3d gravity. Then we focus on the special case of $\Sigma_{0,2}$, so the manifold has topology $S^1\times I\times I$, where one of the intervals ends at asymptotic boundaries and the other ends at EOW branes. We will refer to this as the annulus wormhole, as it is the open analog of the torus wormhole. For this, the computation is done explicitly. Finally, we will provide a derivation for more complicated Riemann surfaces, namely those with negative Euler characteristic. 

Classically, 3d gravity is equivalent to a Chern-Simons theory \cite{Achucarro:1986uwr,Witten:1988hc}. This is expected to extend to the quantum theory perturbatively \cite{Witten:2007kt,Kim:2015qoa}. Non-perturbatively, the two theories are different. Nevertheless, by requiring the gauge fields to correspond to non-degenerate metrics, one picks out the Teichm\"uller component \cite{Krasnov:2005dm,Scarinci:2011np}, and with this restriction, the Chern-Simons theory is equivalent to 3d gravity, except that large diffeomorphisms must be gauged in gravity. For hyperbolic manifolds, the effect of gauging the mapping class group is rather trivial, but the same cannot be said about off-shell manifolds. For example, Virasoro TQFT gives a divergent path integral for the torus wormhole \cite{Collier:2023fwi}, while the direct computation gives a finite answer \cite{Cotler:2020ugk}. It is believed that gauging the mapping class group will resolve this discrepancy \cite{Jafferis:2024jkb}, though it has not been demonstrated explicitly. In this work, we will see that the annulus wormhole is a simple setup where we can see the effect of gauging the mapping class group explicitly, rendering a divergent expression finite.

We will see that the tension of the EOW branes plays an intriguing role. Classically, turning on a nonzero tension modifies the saddle in a controlled way: the partition function is simply multiplied by a factor $g^{\chi}$ \cite{Geng22,Wang:2025bcx}, where $g$ is the BCFT $g$-function, related to the tension via a simple relation \cite{Takayanagi:2011zk}, and $\chi$ is the Euler characteristic of the EOW brane. These factors reproduce the correct factors of $g$ in the BCFT correlators and their statistical moments \cite{Wang:2025bcx}. This topological feature generalizes to finite $c$ for hyperbolic manifolds, where the invariance of this factor under quantum fluctuations follows from the Moore-Seiberg consistency conditions of the open–closed Virasoro TQFT \cite{Jafferis:2025yxt}. For off-shell manifolds, which are the main object of interest in the present work, $g$ plays the important role as a topological coupling that weights different EOW-brane topologies \cite{Jafferis:2025yxt}. From the matrix model or 2d gravity perspective, it is essentially the $\e^{S_0}$ parameter coming from the topological term of JT gravity \cite{Saad:2019lba}; from the purely 3d perspective, where JT or VMS does not appear in the action, we can think of it as the 2d Marolf-Maxfield action \cite{Marolf:2020xie} evaluated on all EOW boundaries.

The paper is organized as follows. In Section~\ref{sec:BCFT}, we review the BCFT partition function on the annulus and use it to formulate the duality between 3d gravity with EOW branes and VMS. 
Sections~\ref{sec:disk}, \ref{sec:cylinder}, and \ref{sec:higher} will be the main sections showing, respectively, that the disk amplitude, the cylinder amplitude, and all other amplitudes of VMS are reproduced by 3d gravity. We then end with comments and open questions in Section~\ref{sec:disc}.

\paragraph{Notations and conventions.} We use $\mathcal{Z}$ for 3d gravity partition functions, $Z$ for BCFT partition functions, and $\mathsf{Z}$ for VMS partition functions. We use $\rho(h)$ for the BCFT spectral density, and $\varrho(E)$ for the VMS matrix model spectrum. The Lorentzian action will be denoted with $S$, while the Euclidean action will be denoted with $I$. Boldface letters such as $\mathbf{P}$ and $\mathbf{m}$ will be $(3g-3+n)$-dimensional vectors, $g$ and $n$ being the genus and the number of boundaries of the Riemann surface $\Sigma_{g,n}$, while symbols like $\vec{P}$ will be used to mean $n$-dimensional vectors.  We will frequently use the Liouville notations
\begin{align}\label{eq:notationsLiou}
 h&=\frac{Q^2}{4}+P^2, \quad Q=b+b^{-1},\quad c=1+6 Q^2.
\end{align}
We also define the following two expressions for the Cardy density:
\begin{align}\label{eq:cardy}
\rho_0(P)&\equiv4\sqrt{2}\,\sinh (2 \pi b P) \sinh (2 \pi b^{-1} P) 
,\\
\label{eq:cardyh}
\rho_0(h)&=
\frac{\rho_0(P)}{2P}.
\end{align}

\section{Duality via BCFT}\label{sec:BCFT}

The central object underlying the duality is the BCFT partition function on the annulus, which is also referred to as the open-string one-loop amplitude. It is defined to be the trace over the open-string Hilbert space: 
\begin{align}\label{eq:trBCFT}
    Z^{(ab)}(\tau)=\tr_{\mathcal{H}^{(ab)}}\!\left(\e^{-2\pi\,\Im(\tau) H^{(ab)}}\right)~~=\quad \vcenter{\hbox{\includegraphics[height=4cm]{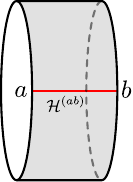}}}~,
\end{align}
where $\mathcal{H}^{(ab)}$ is the Hilbert space of the open-string states, $H^{(ab)}$ is the Hamiltonian generating the direction along the circle, and $2\pi\,\Im(\tau)=\beta$ is the inverse temperature. The geometric result of taking the trace is an annulus, generally with different boundary conditions on the two circular borders, $a$ and $b$. The Hilbert spaces with different pairs of boundary conditions are independent.

Each boundary condition is associated with a function called the $g$-function, defined to be the empty disk partition function with that boundary condition imposed at the circular border of the disk, i.e.,
\begin{align}\label{eq:gdef}
    g_a \equiv Z_{\mathrm{disk}}^{(a)}.
\end{align}

Unlike the torus, the annulus has only one real modulus, which is the ratio between the circle length and the interval length. In the complex plane, it can be represented as the following shaded rectangular region:
\begin{align}\label{eq:moduli}
    \vcenter{\hbox{\includegraphics[height=5cm]{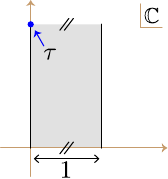}}}~,   
\end{align}
where $\tau$ is a complex number defining the location of the top-left corner of the region. We identify the top and bottom edges to form the annulus, while the left and right edges are genuine borders of the Riemann surface. For the annulus, $\tau$ is purely imaginary. The spatial coordinate ranges from $0$ to $\pi$, so the proper length of the interval is $\pi$ \cite{Cardy:1989ir}; the thermal coordinate ranges from $0$ to $2\pi$, so the proper length of the thermal circle is $2\pi\,\Im(\tau)=\beta$.

For a BCFT, the partition function on the annulus has another interpretation. If we foliate the annulus the other way, i.e., let the circle be the spatial direction and let Euclidean time run along the interval, the partition function is then equal to \cite{Cardy:1989ir}
\begin{align}\label{eq:Ztilde}
    \tilde{Z}^{(ab)}(\tau) \equiv 
    \langle B^{a}|\e^{-\pi \,\Im(\tau)\tilde{H}}|B^{b} \rangle, 
\end{align}
where $B^a$ and $B^b$ are boundary states at the circular borders, with $a$ and $b$ labeling the boundary conditions, and $\tilde{H}$ is the Hamiltonian generating translation in the interval direction. This is often called the closed-string channel. 

One of the BCFT bootstrap conditions is called the open-closed duality \cite{Cardy:1989ir,Cardy:1991tv}, stating that
\begin{align}\label{eq:open-closed}
    \tilde{Z}^{(ab)}(-1/\tau)=Z^{(ab)}(\tau).
\end{align}
The central object underlying the duality will be the RHS, but we will often use the LHS to compute it.

One way to state the conjectured duality is the following. First, denote by $Z^{\rm primary}_{\Sigma_{g,n}\times I}$ the gravitational path integral on $\Sigma_{g,n}\times I$\footnote{We will abuse the notation a little by using $\Sigma_{g,n}$ to denote the Riemann surface with either $n$ punctures/geodesic circles or $n$ asymptotic boundaries.} with EOW branes at the ends of the interval, each of the $n$ asymptotic boundaries being an annulus with moduli
\begin{align}
    \tau\equiv\i t=\i\beta/2\pi.
\end{align} 
We then remove descendant contributions by making the replacement
\begin{align}\label{eq:strip}
    \eta(\i t_i) ={q}_i^{1/24}\prod_{m=1}^{\infty}(1-{q}_i^m)~~\longrightarrow~~ {q}^{1/24}_i,
    \quad 
    {q_i} = \e^{-2\pi t_i},
\end{align}
or, since $\eta$ always appears in the denominator,
\begin{align}\label{eq:stripeq}
    \mathcal{Z}^{(ab)\text{primary}}_{\Sigma_{g,n}\times I}(\beta_1,\dots,\beta_n)
    =
    \mathcal{Z}^{(ab)}_{\Sigma_{g,n}\times I}(\beta_1,\dots,\beta_n) \prod_{i=1}^n\prod_{m=1}^{\infty}(1-{q}_i^m).
\end{align}
Next, expand this function in the energy/conformal weight basis: 
\begin{align}\label{eq:dualZitoVMS}
    \mathcal{Z}^{(ab)\text{primary}}_{\Sigma_{g,n}\times I}(\beta_1,\dots,\beta_n)
    &\equiv 
    \int_{\frac{c-1}{24}}^{\infty}{\d h_1^{(ab)}} \e^{-\beta_1 H_1^{(ab)}}
    \cdots
    \int_{\frac{c-1}{24}}^{\infty}{\d h_n^{(ab)}} \e^{-\beta_n H_n^{(ab)}}
    \nn
    &\times \langle \rho^{(ab)}(h_1)\cdots \rho^{(ab)}(h_n)\rangle_g,
\end{align}
where
\begin{align}\label{eq:Ham}
    H_i^{(ab)}=h^{(ab)}_i-c/24
\end{align}
is the eigenvalue of the Hamiltonian for the state with conformal weight $h_i^{(ab)}$, the shift $-c/24$ being the Casimir energy. 

In terms of inverse Laplace transforms,
\begin{align}\label{eq:dualVMSitoZ}
    \langle \rho^{(ab)}(h_1)\cdots \rho^{(ab)}(h_n)\rangle_g
    =
    \int_{\gamma-\i\infty}^{\gamma+\i\infty}\frac{\d\beta_1}{2\pi\i} \e^{\beta_1 H^{(ab)}_1}
    \cdots
    \frac{\d\beta_n}{2\pi\i} \e^{\beta_n H^{(ab)}_n}
    \mathcal{Z}^{(ab)\text{primary}}_{\Sigma_{g,n}\times I}(\beta_1,\dots,\beta_n),
\end{align}
where $\gamma$ is chosen such that the contour is to the right of the singularities in the complex $\beta$ plane.

The statement of the duality is that
\begin{align}\label{eq:maindual}
\boxed{
    \langle \rho^{(ab)}(h_1)\cdots \rho^{(ab)}(h_n)\rangle_g
    =
    \langle \varrho(E_1)\cdots \varrho(E_n)\rangle^{\rm VMS}_g|_{E_i=h_i-(c-1)/24},\quad \e^{S_0}=g_ag_b,}
\end{align}
where $\langle \varrho(E_1)\cdots \varrho(E_n)\rangle^{\rm VMS}_g$ is the genus-$g$ contribution to the connected $n$-point spectral correlator of the VMS matrix model, with the $\e^{S_0}$ dependence kept. It is related to the VMS partition function correlators $\mathsf{Z}_{g, n}$ via
\begin{align}
    \mathsf{Z}_{g, n}(\beta_1, \ldots, \beta_n)=\int_0^\infty
    {\d E_1}\, \e^{-\beta_1 E_1}
    \cdots 
    {\d E_n}\, \e^{-\beta_n E_n}
    \langle \varrho(E_1)\cdots \varrho(E_n)\rangle^{\rm VMS}_g.
\end{align}
We should emphasize that the angular brackets in \eqref{eq:maindual} have different meanings, despite notational similarity. For the BCFT correlators, it represents gravitational ensemble averaging, while for VMS, it represents quantum correlators of the matrix integral.

From here on, we will often suppress the boundary condition superscripts to simplify the presentation.

\section{Disk amplitude}\label{sec:disk}

The disk amplitude corresponds to $\Sigma_{0,1}$, so the corresponding 3d manifold is a ``thickened disk'', $D\times I$. 
Like in JT gravity, the manifold corresponding to the disk amplitude admits an on-shell geometry. It can be easily obtained by taking a $\mathbb{Z}_2$ quotient of the Euclidean BTZ:
\begin{align}\label{eq:BTZcut}
    \vcenter{\hbox{\includegraphics[width=4cm]{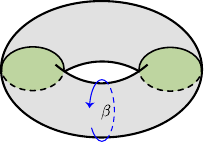}}}~/\mathbb{Z}_2\quad =\quad  \vcenter{\hbox{\includegraphics[width=4cm]{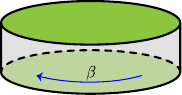}}}~.
\end{align}
Here, the green disks are EOW branes, and $\beta$ labels the thermal circle. 

From the BCFT perspective, this computes the annulus partition function. The simplest way to obtain the full finite-$c$ partition function on this manifold is to read it off from the open-closed version of Virasoro TQFT \cite{Jafferis:2025yxt}, built from the BCFT version \cite{Cardy:1991tv,Lewellen:1991tb} of Moore-Seiberg consistency conditions \cite{Moore:1988uz,Moore:1988qv}. The expression in the $h$ basis is given by
\begin{align}\label{eq:diskdensity}
\langle\rho^{(ab)}(h)\rangle_0=g_ag_b\,\rho_0(h),
\end{align}
where $\rho_0(h)$ is the Cardy density given in \eqref{eq:cardyh} and the $g$-functions are defined via \eqref{eq:gdef}.
The particular TQFT condition that is needed to derive this is the open-closed duality \eqref{eq:open-closed}, uplifted to 3d. (CFT or BCFT bootstrap relations have a natural uplift to 3d handlebodies \cite{Witten:1988hf,Collier:2023fwi,Jafferis:2024jkb,Jafferis:2025yxt}.)

To be explicit, we can start with the expression in the closed-string channel,
\begin{align}
    \langle\tilde{Z}(\tau)\rangle=g_ag_b\,\chi_0(\tau),
\end{align}
which is the vacuum character multiplied by $g$-functions. We then use \eqref{eq:open-closed} to write the open-channel expression as
\begin{align}
    \mathcal{Z}_{D\times I}(\tau)&=\langle Z(\i t)\rangle=\langle\tilde{Z}(\i/t)\rangle
    \nn
    &=
    g_ag_b\int_0^{\infty} \d P\, \mathbb{S}_{\mathbb{1}P}[\mathbb{1}]\,\chi_P(\i t)
=
    \int_0^{\infty}\d P \,g_ag_b\rho_0(P)\frac{q^{P^2}}{\eta(\i t)},
\end{align}
where $\mathbb{S}_{\mathbb{1}P}[\mathbb{1}]=\rho_0(P)$ is the modular crossing kernel. Using \eqref{eq:notationsLiou}, \eqref{eq:stripeq}, and \eqref{eq:Ham},
\begin{align}
   \mathcal{Z}^{\rm primary}_{D\times I}(\i t) &=
   \int_0^{\infty}\d P \,\rho_0(P)\frac{q^{P^2}}{q^{1/24}}
   =\int_{\frac{c-1}{24}}^{\infty}\d h \,\rho_0(h)\,{\e^{-\beta H}}.
\end{align}    
From \eqref{eq:dualZitoVMS}, we can read off that the disk amplitude is indeed given by \eqref{eq:diskdensity}, with support $h>(c-1)/24$.

\section{Cylinder amplitude}\label{sec:cylinder}

We now compute explicitly the matrix cylinder amplitude, corresponding to
\begin{align}
    \mathcal{Z}_{A\times I}(\i t_1,\i t_2)=\langle Z(\i t_1)Z(\i t_2)\rangle_0,
\end{align}
where $A$ stands for the annulus or cylinder, $Z(\i t_i)$ is the open-channel BCFT annulus partition function defined in \eqref{eq:trBCFT}, and the subscript $0$ means the leading (genus-0) contribution. This is a two-sided wormhole connecting two annular asymptotic boundaries, so we call it the annulus wormhole. 

We will follow closely the computation of \cite{Cotler:2020ugk}, but it will differ at various places. In particular, we do not need to perform an analytic continuation of the symplectic form or the volume form, and we carry out an explicit treatment of the mapping class group.

To begin with, the Lorentzian action of Einstein gravity in Chern-Simons formalism is given by
\begin{align}\label{eq:actionCS}
S=-\frac{k}{4 \pi} \int \operatorname{tr}\left[A \wedge \d A+\frac{2}{3} A \wedge A \wedge A\right] - (A\to \bar{A})+S_{\rm bdy}+S_{\rm EOW},\quad k=\frac{1}{4G},
\end{align}
where
\begin{align}\label{eq:actionasymp}
S_{\text{bdy}}=S_{\text{EH}}^{\partial}+S_{\rm GHY}+S_{\rm counter}
\end{align}
is the action at asymptotic boundaries, composed of the total derivative part of the Einstein-Hilbert action, the Gibbons-Hawking-York action, and the counterterm, and $S_{\rm EOW}$ is the action for the EOW branes
\cite{Karch:2000ct,Karch:2000gx,Takayanagi:2011zk,Fujita:2011fp}. We will only need the Euclidean version of the EOW action, which is given by 
\begin{align}\label{eq:EOWaction}
I_{\text{EOW}}=I_{\text{EH}}^{\partial}-\frac{1}{8 \pi G_N} \int \sqrt{h}\, K -\chi \log g.
\end{align}
Here, instead of a tension term, our model includes a topological term that depends on the $g$-function. It is equivalent to having a tension term for on-shell manifolds \cite{Geng22,Wang:2025bcx}. The form of the second term in terms of the gauge fields can be found in \cite{Takayanagi:2020njm}. 

We define generators of the fundamental representation of $\SL$ via\footnote{Explicit forms are given by (denoting the Pauli matrices with $\sigma_A$)
\begin{align}\label{eq:Jconvention}
J_0=-\i \sigma_2/2, \quad J_1= \sigma_1/2, \quad J_2=\sigma_3/2.
\end{align}}
\begin{align}
\left[J_A, J_B\right]=\varepsilon_{A B C} J^C, \quad \operatorname{tr}\left(J_A J_B\right)=\frac{1}{2} \eta_{A B},\quad A,B\in\{0,1,2\},
\end{align}
where $\varepsilon_{012}=-1$ (this convention also affects the overall sign in \eqref{eq:actionCS}), and the Minkowski metric $\eta_{AB}$ is used to raise and lower indices. Similarly for $\bar{J}_A$.

As for the boundary conditions at EOW branes, we impose Neumann boundary conditions. This translates to \cite{Takayanagi:2020njm}
\begin{align}\label{eq:BCgen}
P_{AB} (A+\bar{A})_i^B+2\varepsilon_{ABC} n^B \partial_i n^C=0,\quad 
 n_A (A-\bar{A})_i^A=0,
\end{align}
where $i$ denotes a spacetime index tangent to the brane, $n^A$ is a unit timelike vector related to the spacetime normal vector to the brane $N^\mu$ and the dreibein $e^{A}_{\mu}$ via $n^A=e^{A}_{\mu}N^\mu$, and $P_{AB}=\eta_{AB}+n_An_B$ is the projector onto the plane orthogonal to $n^A$.

To get started, let us set up the coordinate system as follows:
\begin{align}\label{eq:annuworm}
    \vcenter{\hbox{\includegraphics[height=4cm]{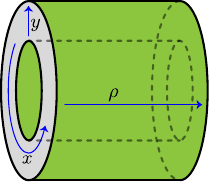}}}
    \quad = \quad 
    \vcenter{\hbox{\includegraphics[height=4cm]{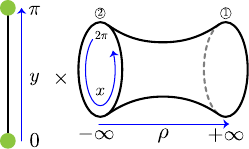}}}
    ~.
\end{align}
Here, $x$ is an angular coordinate, $x\sim x+2\pi$; $y \in [0,\pi]$ is the interval direction with EOW branes at both ends (shown in green); $\rho$ runs from $-\infty$ (boundary 2) to $+\infty$ (boundary 1). The shape of the diagram on the right is a reminder that the metric grows near the asymptotic boundaries. 

The moduli of the boundary annuli are captured by the asymptotic metric. Near boundary 1, 
\be 
\d s^2 = \frac{\e^{2\rho}}{4}(\d y^2 + t_1^2 \,\d x^2) + \d\rho^2 +  \O(\e^{-\rho}),
\ee
where $t_1$ is the real modulus of boundary 1. The expression for boundary 2 is similar. While we can proceed with this choice, which would directly compute for us $\langle Z(\i t_1)Z(\i t_2)\rangle_0$, in order to follow the notation of \cite{Cotler:2020ugk} as closely as possible, we will instead pick a different conformal frame such that
\begin{align}
    \d s^2 &= \frac{\e^{2\rho}}{4}\left(\d x+\frac{\d y}{t_1^2}\right)^2 + \d\rho^2 +  \O(\e^{-\rho})\\
    &= \frac{\e^{2\rho}}{4}\left|\d x+\tau_1 \d y\right|^2 + \d\rho^2 +  \O(\e^{-\rho}).
\end{align}
This would instead compute $\langle \tilde{Z}(\tau_1)\tilde{Z}(\tau_2) \rangle_0$, but it is related to what we are interested in via \eqref{eq:open-closed}.

In terms of gauge connections in the Chern-Simons formulation, this translates to
\begin{equation}\label{eq:Aasymp1}
    A = \frac{1}{2} \begin{bmatrix}
        \d\rho & 0 \\
        \e^{\rho}(\d x + \bar{\tau}_1\d y) & -\d\rho
    \end{bmatrix}
    + \O(\e^{-\rho}),
    ~~
    \bar{A} = \frac{1}{2} \begin{bmatrix}
        {-}\d\rho & {-}\e^{\rho}(\d x + \tau_1\d y) \\
        0 & {+}\d\rho
    \end{bmatrix}
    + \O(\e^{-\rho}).
\end{equation} 
Similarly, near boundary 2, 
\begin{equation}\label{eq:Aasymp2}
    A = \frac{1}{2} \begin{bmatrix}
        \d\rho & \e^{-\rho}(\d x + \bar{\tau}_2\d y) \\
        0 & -\d\rho
    \end{bmatrix}
    +\O(\e^{\rho}),
    ~~
    \bar{A} = \frac{1}{2} \begin{bmatrix}
        -\d\rho & 0 \\
        -\e^{\rho}(\d x + \tau_2\d y) & \d\rho
    \end{bmatrix}
    + \O(\e^{\rho}).
\end{equation}
These expressions are actually identical to those for the torus wormhole, but the coordinates have different ranges, distinguishing the annuli from the tori.  

We then Euclideanize the action, which in our coordinates becomes
\begin{align}\label{eq:Itot}
    I &= I_{\rm bulk}+I_{\rm bdy},\\
    I_{\rm bulk} &= \frac{\i k}{4\pi}\int \d^3x\, \tr\big(A_\rho\del_y A_x - A_x \del_y A_\rho + 2 A_y F_{x\rho}\big) - (A\rightarrow \bar{A}),
    \\
    I_{\rm bdy} &= \frac{\i k}{4\pi} \int_{1} \d^2 x \, \tr \big(\bar{\tau}_1A_x^2 -\tau_1 \bar{A}_x^2\big)
    + \frac{\i k}{4\pi}\int_{2} \d^2 x \,\tr\big(\bar{\tau}_2A_x^2 - \tau_2\bar{A}_x^2\big).
\end{align}
Note that only the asymptotic boundaries have non-vanishing action. The topological term in the EOW brane action vanishes in this case because the Euler characteristic for the annulus is zero, and the other terms vanish due to the boundary conditions imposed at the EOW branes. 

Noticing that $A_y$ and $\bar{A}_y$ appear as Lagrange multipliers in the action, we path integrate them first, setting the spatial field strength $F_{x\rho}= \bar{F}_{x\rho}=0$.

The flatness conditions means that the connections can be written as
\be 
A_i = \tilde{G}^{-1}\del_i\tilde{G},\quad \bar{A}_i = G_0^{-1}\tilde{\bar{G}}^{-1}\del_i\tilde{\bar{G}}G_0,\quad i\in\{x,\rho\}.
\ee 
where $\tilde{G}$ and $\tilde{\bar{G}}$ are elements of the $\SL$ gauge group. The $G_0$ is a convenient choice in the definition of $\tilde{\bar{G}}$ to simplify the discussion later. 
At the EOW branes, the boundary conditions imply that $\tilde{G}$ and $\tilde{\bar{G}}$ are $\SL$ gauge equivalent at the brane, i.e., $\tilde{\bar{G}}= \e^{g(y)}\tilde{G}$ for some $g(y)$.

The group-valued functions $\tilde{G}$ and $\tilde{\bar{G}}$ can be multi-valued around the $x$ cycle, which is related to the gauge holonomy around the circle. Following the notation of \cite{Cotler:2020ugk}, we characterize this by 
\be \label{eq:GtilandG}
\tilde{G} = \e^{\l(y)x}G,\quad
\tilde{\bar{G}} = \e^{\bar{\l}(y)x}\bar{G},
\ee 
with $G$ and $\bar{G}$ single-valued. In terms of $G$, 
\begin{align}
    A_x = \tilde{G}^{-1}\del_x \tilde{G} = G^{-1}\l(y) G + G^{-1}\del_x G.
\end{align}

The reparameterization \eqref{eq:GtilandG} has introduced a new redundancy
\be 
\tilde{G} \sim \e^{h(y)}\tilde{G},\quad
\tilde{\bar{G}} \sim \e^{\bar{h}(y)}\tilde{\bar{G}},
\ee 
which is partially fixed by setting 
\be 
\tilde{G} = \e^{b(y)xJ_1}G,\quad
\tilde{\bar{G}} = \e^{\bar{b}(y)xJ_1}\bar{G}.
\ee 
Here, $J_1$ is one of the generators in the fundamental representation, defined in \eqref{eq:Jconvention}. Only allowing such hyperbolic conjugacy class corresponds to picking the correct component of the phase space of the Chern-Simons theory that Virasoro TQFT restricts to.

Next, we further decompose $G$ and $\bar{G}$ as
\be 
G = \e^{\phi J_1} \e^{\Lambda J_2} \e^{\psi(J_1 - J_0)}
,\quad
\bar{G} = \e^{\bar{\phi}J_1} \e^{-\bar{\Lambda}J_2} \e^{\bar{\psi}(J_1 + J_0)},
\ee 
and denote 
\be \label{eq:bigsmallphi}
\Phi = b(y)x + \phi,\quad \bar{\Phi} = \bar{b}(y)x + \bar{\phi},
\ee 
with residual $U(1)$ redundancies
\be 
\phi(x, y, \rho) \sim \phi + a(y),\quad
\bar{\phi}(x, y, \rho) \sim \bar{\phi} + \bar{a}(y).
\ee 
In this gauge fixing, $\tilde{G}$ and $\tilde{\bar{G}}$ are now $U(1)$ gauge equivalent at the EOW branes instead of $\SL$. In other words,
\be 
\Phi(y) = \bar{\Phi}(y) + g(y)\big|_{\rm EOW},
\ee 
for some $g(y)$, or equivalently,
\be 
b= \bar{b}|_{\rm EOW}, \quad \phi= \bar{\phi}+ g|_{\rm EOW}.
\ee 
However, $\phi$ and $\bar{\phi}$ already have this redundancy, so the EOW brane boundary conditions do not constrain $\phi$. It only enforces that $b$ and $\bar{b}$ are equal at the EOW branes. 

In terms of these new functions, the connections now take the form 
\begin{align} 
A &= \frac{1}{2}\begin{bmatrix} 
\d\Lambda - \psi \e^\Lambda \d\Phi \ \ & 2(\d\psi + \psi \d\Lambda) + (\e^{-\Lambda} - \psi^2 \e^\Lambda){\d\Phi}\\
 \e^\Lambda \d\Phi & -\d\Lambda + \psi \e^\Lambda \d\Phi
\end{bmatrix},
\\
\bar{A} &=\frac{1}{2} \begin{bmatrix} 
-\d\bar{\Lambda} + \bpsi \e^{\bar{\Lambda}} \d\bar{\Phi}  &  \e^{\bar{\Lambda}} \d\bar{\Phi} \\
2(\d\bpsi + \bpsi \d\bar{\Lambda}) + (\e^{-\bar{\Lambda}} - \bpsi^2 \e^{\bar{\Lambda}}){\d\bar{\Phi}} \ \ & \d\bar{\Lambda} - \bpsi \e^{\bar{\Lambda}} \d\bar{\Phi}
\end{bmatrix}.
\end{align}
Setting the asymptotic boundary conditions \eqref{eq:Aasymp1} and \eqref{eq:Aasymp2}, the fields are related near the boundaries by 
\begin{align}\label{eq:bdyrelations}
\Lambda \approx \ln \left(\frac{\e^\rho}{\Phi^{\prime}}\right), \quad \psi \approx-\frac{\e^{-\rho} \Phi^{\prime \prime}}{\Phi^{\prime}}, \quad \bar{\Lambda} \approx \ln \left(\frac{\e^\rho}{\bar{\Phi}^{\prime}}\right), \quad \bar{\psi} \approx-\frac{\e^{-\rho} \bar{\Phi}^{\prime \prime}}{\bar{\Phi}^{\prime}}.
\end{align}

Now, let us explain the key fact that will simplify the calculation. In analogy with the CFT doubling trick of Cardy \cite{Cardy:1984bb}, which has a natural uplift to Chern-Simons theory, we perform the following actions, as illustrated by the diagrams (showing only the $y$ direction):
\begin{align}\label{eq:doub}
    \vcenter{\hbox{\includegraphics[height=3.5cm]{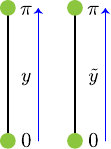}}}
    \quad \longrightarrow \quad 
    \vcenter{\hbox{\includegraphics[height=3.5cm]{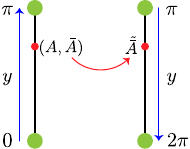}}}
    \quad \longrightarrow \quad 
    \vcenter{\hbox{\includegraphics[height=3.5cm]{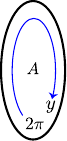}}}
    ~.
\end{align}
The idea is to relocate $\bar{A}$ onto a copy of the manifold in a way such that the information about $A$ and $\bar{A}$ on the original manifold can be repackaged into a single field $A$ on the two copies, which are then joined to become a single manifold, which we call the doubled manifold. In general, this is done by designing the map in a way so that the resulting configuration is describable by a chiral action. We now demonstrate this in a simple gauge where $n^A$ is chosen to be $(1,0,0)$. The boundary conditions \eqref{eq:BCgen} then simplify and can be packaged into a single equation as
\begin{align}\label{eq:BCgauged}
    \mathcal{A}=G_0\bar{\mathcal{A}}G_0^{-1},\quad G_0=\begin{bmatrix}
        0 & 1 \\
        -1 & 0
    \end{bmatrix}.
\end{align}
In particular, this gauge choice is compatible with the choice at asymptotic boundaries, which we specify later in \eqref{eq:Aasymp1} and \eqref{eq:Aasymp2}.

Let us now describe the doubling trick in detail in this gauge. First of all, make a copy of the manifold and label it with coordinates $(\tilde{\rho},\tilde{x},\tilde{y})$. These coordinates take the same range of values as $(\rho,x,y)$. Next, relocate the $\bar{A}$ field to the second copy, by which we mean defining $\tilde{\bar{A}}(\tilde{\rho},\tilde{x},\tilde{y})\equiv \bar{A}(\rho=\tilde{\rho},x=\tilde{x},y=\tilde{y})$. Now, with an eye toward what follows, let us define $\tilde{{A}}$ such that 
$\tilde{{A}}_{\tilde{y}}(\tilde{\rho},\tilde{x},\tilde{y})
=\tilde{\bar{A}}_{\tilde{y}}(\tilde{\rho},\tilde{x},\tilde{y})$
and $\tilde{{A}}_i(\tilde{\rho},\tilde{x},\tilde{y})
=G_0^{-1}\tilde{\bar{A}}_i(\tilde{\rho},\tilde{x},\tilde{y})G_0$, $i\in\{\tilde{x}, \tilde{\rho}\}$. Then we do the ``doubling" by combining the two manifolds into a single one by letting $y\equiv 2\pi-\tilde{y}$, $\tilde{y}\in[0,\pi]$, extending the range of $y$ to $[0,2\pi]$, as well as identifying $x=\tilde{x}$ and $\rho=\tilde{\rho}$. This sets $\tilde{A}_y(\rho,x,y)=-\tilde{A}_{\tilde{y}}(\rho,x,\tilde{y}=2\pi-y)$, $y\in[\pi,2\pi]$, where the minus sign is due to the Jacobian, while $\tilde{A}_i(\rho,x,y)=\tilde{A}_i(\rho,x,\tilde{y}=2\pi-y)$. Finally, noticing that $A$ has only been defined for $y\in[0,\pi]$ (the original manifold), we are free to extend the definition of $A(\rho,x,y)$ by setting it to $A(\rho,x,y)=\tilde{A}(\rho,x,y)$, $y\in[\pi,2\pi]$, so all information has been packaged into a single field $A$ on the doubled manifold. To summarize, what this procedure achieves is to flip the sign of $\bar{A}_{y}$ and to gauge transform $\bar{A}_i$ before sending them to an image location in the doubled manifold. We should emphasize that this is a mere redefinition so far.

The non-trivial aspects of this redefinition have to do with the EOW boundary conditions and the action. By \eqref{eq:BCgauged}, the components of the field $A$ that is parallel to the brane is continuous across $y=0,\pi$, so it remains a continuous configuration in the doubled manifold; on the other hand, $A_y$ is not necessarily continuous, but this does not lead to any divergence, owing to the fact that it only appears as a Lagrange multiplier in the action.

To see the consequence of the doubling trick on the action, notice that the $\bar{A}$ part of the bulk action in \eqref{eq:Itot} flips sign under this redefinition, while the boundary action remains invariant. So we can simply replace the action \eqref{eq:Itot} with
\begin{align}\label{eq:Ichiral}
    I_{\rm bulk}&=\frac{\i k}{4\pi}\int_{A\times S^1} \d^3x\, \tr\big(A_\rho\del_y A_x - A_x \del_y A_\rho + 2 A_y F_{x\rho}\big),\\
    I_{\rm bdy} &= -\frac{\i k}{4\pi} \int_{1} \d^2 x \, \tr \big({\tau}_1A_x^2 \big)
    + (1\to2),
\end{align}
where we have used the fact that $\tau_i$ is purely imaginary for us, and the labels 1 and 2 now refer to the two asymptotic boundaries of the doubled manifold (each of topology $I\times S^1$). This has turned the problem into one looking like the torus case, but with only a chiral half of the action. The problem should now be more familiar to us, as this is just the action of chiral 3d gravity \cite{Li:2008dq}. We are free to confuse the two theories as far as the action is concerned, though as we shall see, it is important that we are working with 3d gravity with EOW branes, not chiral 3d gravity, when discussing global issues such as the mapping class group. 

In terms of the fields defined earlier, the action is now given by
\begin{align}
    {I}_{\text{bulk}} =\frac{\i k}{8\pi} \int \d^3x \Big[ &-2\del_\rho(\e^\La\del_x\Phi\del_y\psi) + 2\del_x(\e^\La\del_\rho\Phi\del_y\psi) - \del_y(\e^\La(\del_\rho\Phi\del_x\psi - \del_x\Phi\del_\rho\psi))\nonumber\\
    &+ \del_x(\del_y\La\del_\rho\La) - \del_\rho(\del_y\La \del_x\La) + \del_x(\del_y\Phi\del_\rho\Phi) - \del_\rho(\del_y\Phi \del_x\Phi)\Big],
\end{align}
where we have dropped the label under the integral sign here, and it should henceforth be understood that all $y$ integrals will range between $0$ and $2\pi$ unless otherwise stated.

As $x$ is a compact direction, the total $x$-derivative terms would vanish for periodic fields. (The only field non-periodic in $x$ here is $\Phi = \phi + b(y)x$.) This expression therefore simplifies significantly, in fact reducing to a purely boundary term (presenting only the contribution at boundary 1):
\begin{align}
{I}_{\text{bulk},1} =-\frac{\i k}{8\pi} \int_{1} \d^2x  \Big(2\e^\La\del_x\Phi\del_y\psi + \del_y\La\del_x\La + \del_y\Phi\del_x\Phi  -\del_yb\,\del_x(x\phi)\Big).
\end{align}
The boundary action is (again presenting only for boundary 1)
\be 
\begin{split}
    I_{\text{bdy},1} 
    &= -\frac{\i k}{8\pi} 
    \int_{1} \d^2 x 
    \, {\tau}_1\Big( (\del_x\La)^2 + (\del_x\Phi)^2 + 2\e^\La\del_x\Phi\del_x\psi \Big).
\end{split}
\ee
Substituting the boundary relations \eqref{eq:bdyrelations}, the total action at boundary 1 becomes
\begin{align}
    I_1 
    = &-\frac{\i k}{4\pi} \int_1 \d^2 x\Bigg[ \frac{1}{2}\frac{\Phi''(\del_y +{\tau}_1\del_x)\Phi'}{\Phi'^2} -(\del_y +{\tau}_1\del_x)\!\left(\frac{\Phi''}{\Phi'}\right)  + \inv{2}\Phi'(\del_y +{\tau}_1\del_x)\Phi \Bigg]
    \nonumber\\
    &+ \frac{\i k}{8\pi}\int_1 \d^2 x\,\dot{b} \,\del_x(x\phi),
\end{align}
where the prime means $\del_x$ and the dot means $\d/\d y$. 
Dropping total derivatives of periodic functions and with a bit of rewriting,
\be
    I_1 
    =
    \frac{k}{4\pi} \int_1 \d^2x \Bigg[\frac{\Phi_1''\del_1\Phi_1'}{\Phi_1'^2} 
    -
    \frac{\i}{2}({\tau}_1\Phi_1'^2 + \phi_1' \dot{\phi}_1) \Bigg]
     +\frac{\i k}{4} \int_1 \d y \Big[ \phi_{1,0}\dot{b} - b\dot{\phi}_{1,0} \Big],
\ee 
where $\Phi_1 \equiv \lim_{\rho\to\infty} \Phi$,  
$\del_1 \equiv -\frac{\i}{2}({\tau}_1\del_x+\del_y)$, 
\begin{align}\label{eq:phizeromode}
    \phi_{i,0}(y) \equiv \inv{2\pi}\int \d x \,\phi_i(x, y),\quad i\in\{1,2\},
\end{align}
are the zero modes in the $x$ direction.

On the other boundary, the bulk action comes with an extra minus sign, while the boundary action comes with the same sign, so the contribution to the action from boundary 2 is given by
\be 
    I_2 
    = \frac{k}{4\pi} \int_2 \d^2x \left[ \frac{\Phi_2''\del_2\Phi_2'}{\Phi_2'^2} -\frac{\i}{2}({\tau}_2\Phi_2'^2 - \phi_2'\dot{\phi}_2)\right] + \frac{\i k}{4}\int_2 \d y
    \Big[ \phi_{2,0}\dot{b} - b\dot{\phi}_{2,0} \Big],
\ee 
where $\Phi_2 \equiv \lim_{\rho\to-\infty} \Phi$ and $\del_2 =- \frac{\i}{2}({\tau}_2 \del_x-\del_y)$. The total action is the sum of $I_1$ and $I_2$.

Let us now analyze the zero-mode part of the action. The terms involving solely $\Phi'_i$ or higher derivatives do not involve the zero modes, so the only source of zero modes is
\be \label{eq:zeromodeaction}
I_0
\equiv \frac{\i k}{4} \int \d y \Big[ \Delta\phi_{0}\dot{b} - b\Delta\dot{\phi}_{0}\Big], \quad \Delta\phi_0\equiv \phi_{1,0}-\phi_{2,0}.
\ee
This tells us that the difference between zero modes on the two boundaries $\Delta\phi_0$ and the holonomy parameter $b$ are conjugate variables.  

Before moving on to the path integral, we note one crucial point where gravity and gauge theories differ: how the mapping class group is gauged. The space of flat connection on the spatial slice (fixed $y$) is characterized by the radial and angular holonomies. In our notation, they are given by
\begin{align}
    H_x& = \mc{P}\oint \d x \exp\big(\!-\! A_x(\rho\to \infty, x, y)\big)  \nonumber\\
    &= G^{-1}(x=2\pi, y, \rho\to\infty)\,G(x=0, y, \rho\to\infty)\nonumber\\ 
    &= \e^{-\psi_1(J_1 - J_0)} \e^{-\La_1 J_2} \e^{-2\pi b J_1}\e^{\La_1 J_2}\e^{\psi_1(J_1 - J_0)},
\end{align}
and
\begin{align}
    H_\rho &= \mathcal{P}\int \d\rho \exp\big(\! -\!A_\rho(\rho, x, y) \big) \nonumber\\
    &= G^{-1}(\rho\to \infty, x, y)\,G(\rho \to -\infty, x, y) \nonumber\\
    &=\e^{-\psi_1(J_1 - J_0)}\e^{-\La_1 J_2} \e^{-\Phi_1 J_1} \e^{\Phi_2 J_1}\e^{\La_2 J_2}\e^{\psi_2(J_1 - J_0)}.
\end{align}
The spatial mapping class group can be represented by
\begin{align}\label{eq:spatial}
    &y\to y, \quad \rho \to \rho,\quad x \to x + f_n(\rho),\nonumber
    \\
    \text{where}\quad &f_n(-\infty) = 0,\quad  f_n(\infty)= 2\pi n,\quad  n \in \mb{Z}.
\end{align}
For example, we can further pick representatives by setting
\be 
f_n(\rho) = \frac{2\pi n \,\e^{\rho}}{\e^{\rho} + \e^{-\rho}}.
\ee 

Since $\del_\rho f_n(\rho)=0$ at $\rho\to\pm\infty$, the mapping class group acts trivially on the values of the gauge connections at the boundaries.
Nevertheless, it has an important effect, as it can change the holonomy. As $H_x$ can be evaluated just at the boundary, it is invariant under the mapping class group action, so we only need to look at $H_\rho$. The holonomy path of $H_\rho$ changes under the mapping class group element as follows (demonstrated for $n=1$):
\begin{align}\label{eq:holonomies}
    \vcenter{\hbox{\includegraphics[height=3.5cm]{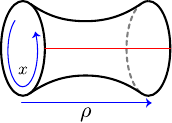}}}
    \quad \longrightarrow \quad 
    \vcenter{\hbox{\includegraphics[height=3.5cm]{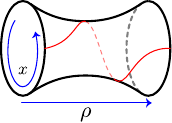}}}
~
.
\end{align}
To compute the new holonomy, notice that we can deform the path into
\begin{align}
    \vcenter{\hbox{\includegraphics[height=3.5cm]{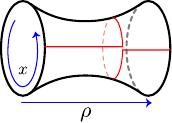}}}~,
\end{align}
which simplifies in the limit the $x$ loop reaches $\rho=\infty$. Therefore, the effect of the mapping class group action is just
\be 
H_\rho ~\to ~H_x H_\rho,
\ee 
where
\begin{align}
    H_\rho &= \e^{-\psi_1(J_1 - J_0)}\e^{-\La_1 J_2} \e^{-\Delta\Phi J_1}\e^{\La_2 J_2}\e^{\psi_2(J_1 - J_0)},\\
    H_x H_\rho &= \e^{-\psi_1(J_1 - J_0)}\e^{-\La_1 J_2} \e^{-(\Delta\Phi + 2\pi b) J_1} \e^{\La_2 J_2}\e^{\psi_2(J_1 - J_0)}.
\end{align}
In gravity, gauging the mapping class group therefore translates to
\begin{align}
    H_\rho \sim H_xH_\rho\implies \Delta\Phi(x, y)\sim \Delta\Phi(x, y)+2\pi b(y),
\end{align}
where $\Delta\Phi \equiv \Phi_1 - \Phi_2$. Using \eqref{eq:bigsmallphi} and \eqref{eq:phizeromode}, this means
\begin{align} \label{eq:Dphiperiod}
\Delta\phi_0(y)\sim \Delta\phi_0(y) + 2\pi b(y).
\end{align}
In conclusion, the effect of gauging the mapping class group is to turn $\Delta\phi_0(y)$ into a compact field. 

We are almost ready to do the zero-mode path integral
\be 
Z_0 \sim \int[\d b(y)][\d\xi(y)] \exp\left(-I_0\right).
\ee
However, it is clear from \eqref{eq:Dphiperiod} that the period of $\phi(y)$ is not constant in $y$. This makes the path integral complicated. Instead, we define
\begin{align}
    Y(y) \equiv \frac{\Delta\phi_0(y)}{b(y)},
\end{align}
which has a constant period of $2\pi$. In terms of $Y(y)$, the zero-mode action \eqref{eq:zeromodeaction} is given by
\be \label{eq:actionb2Y}
    I_{0}  = -\frac{\i k}{4} \int \d y\, b^2(y)\dot{Y}(y).
\ee 
To perform the path integral, we expand the field in Fourier modes,
\be 
b^2(y) = b^2_0 + \sum_{n>0}b^2_n \e^{\i ny} + b^{2*}_n \e^{-\i ny},
\quad 
Y(y) = Y_0 + \sum_{n>0}Y_n \e^{\i ny} + Y_n^*\e^{-\i ny},
\ee
where $b^2_n$ are the Fourier coefficients of $b^2(y)$ (not the squares of the Fourier coefficients of $b(y)$, despite what the notation might suggest).

The periodicity $Y(y)\sim Y(y) + 2\pi$ now translates to the zero-mode periodicity $Y_0\sim Y_0 + 2\pi$, while the nonzero modes $Y_n$ are unrestricted. (Had we Fourier transformed $\Delta \phi_0(y)$ rather than $Y(y)$, the periodicity condition would be very complicated.) With this, the zero-mode action \eqref{eq:actionb2Y} turns into
\be 
I_0= \sum_{n>0} \i kn\pi\big(\Im{(Y_{n})}\Re(b^2_{n})- \Re(Y_{n})\Im(b^2_{n})\big) .
\ee 
Performing the path integral in $Y$, which now becomes a product of the ordinary integrals in $Y_n$, we obtain delta functions $\de(\Re(b^2_{n}))$ and $\de(\Im(b^2_{n}))$ from the $n>0$ modes, setting all Fourier modes of $b^2(y)$ to zero except the zero mode, so
\begin{align}
    b^2(y)=b_0^2=\text{constant}.
\end{align}
The zero-mode path integral
\begin{align}
    Z_0\sim \int [\mathrm{d}b^2][\mathrm{d}Y]\,\e^{-I_0}
\end{align}
therefore simplifies to
\be 
Z_0=\mc{N}\int \d b^2_0\, \d Y_0 = 2\pi\mc{N}\int \d b^2_{0}, 
\ee 
where $\mc{N}$ collects all normalization factors from the integrals and requires a Zeta function regularization to be finite \cite{Cotler:2020ugk}. 

We are now ready to move on to the rest of the path integral. Note that we have integrated over $\Delta\phi_0$, so $\phi_1$ and $\phi_2$ become independent fields without zero modes, i.e.,
\be 
\int \d x \,\phi_1(x,y) = \int \d x\, \phi_2(x,y) =0.
\ee
Even though we are supposed to integrate only over such field configurations, we are free to integrate in an auxiliary zero mode for each of them and treat them as gauge redundancies. More precisely, take each $\phi_i$ to again be a general field, but with identifications
\be 
\phi_1 \sim \phi_1 + a_1(y), \quad \phi_2 \sim \phi_2 + a_2(y).
\ee 

After subtracting off the zero-mode part of the action, what remains is
\begin{align}
    I'\equiv I-I_0
    = \frac{k}{4\pi} \int_1 \d^2x  \Big[\frac{\Phi_1''\del_1\Phi_1'}{\Phi_1'^2} + \Phi_1'\del_1\Phi_1 \Big]+(1\to2).
\end{align}
To proceed to the next step, recalling that $b= b_0$ is a constant as a result of the zero-mode path integration, we can perform a field redefinition
\be 
\Phi_i = bx + \phi_i \to b\phi_i,
\ee 
turning the action into two copies of the Alekseev-Shatashvili theory \cite{Alekseev:1988ce}:
\begin{align}
    I'=
    \frac{k}{4\pi} \int_1 \d^2x  \left[\frac{\phi_1''\del_1\phi_1'}{\phi_1'^2} + b^2\phi_1'\del_1\phi_1 \right]+(1\to2)=I_{\rm AS}[\phi_1;\tau_1,b]+I_{\rm AS}[\phi_2;\tau_2,b].
\end{align}
The path integral is now simply
\be 
\langle \tilde{Z}(\tau_1)\tilde{Z}(\tau_2) \rangle_0 = 2\pi\mc{N}\int \d b^2 Z_{\rm AS}(\tau_1|b)Z_{\rm AS}(\tau_2|b).
\ee 

The partition function for the Alekseev-Shatashvili theory has been computed and is one-loop exact \cite{Cotler:2018zff}:
\be 
Z_{\rm AS}(\tau|b) = {q}^{h-\frac{c}{24}}\prod_{n=1}^{\infty}\inv{1-{q}^n},\quad h = \frac{c-1}{24}(1 + b^2),\quad {q} = \e^{2\pi \i\tau}, \quad c = 6k+1.
\ee 
Performing the integral over $b^2$ (via a change of variable to $h$) and using \eqref{eq:open-closed}, we get
\begin{align}
    \mathcal{Z}_{A\times I}(\i t_1,\i t_2)
    &=\langle {Z}(\i t_1){Z}(\i t_2) \rangle_0
    =
    \langle \tilde{Z}(\i/t_1)\tilde{Z}(\i/t_1) \rangle_0 
    \nn&= \frac{\mc{N}}{\eta(\i/t_1)\eta(\i/t_2)}\int_{\frac{c-1}{24}}^{\infty} \d h\,  \e^{-2\pi \left(h-\frac{c{-1}}{24}\right)\left(\inv{t_1} + \inv{t_2}\right)}\nonumber\\
    &= \frac{\mc{N}}{\eta(\i t_1)\eta(\i t_2)}\frac{\sqrt{t_1t_2}}{2\pi(t_1 + t_2)},   
\end{align}
where $\mathcal{N}$ has been redefined to absorb some extra constant factors, and we have used the property that $\eta$ transforms like  $\eta(-1/\tau)=\sqrt{-\i \tau}\,\eta(\tau)$ under the modular S transformation. 

Finally, we use \eqref{eq:strip} to strip off the contributions from the descendants and use \eqref{eq:dualVMSitoZ} to obtain the cylinder amplitude:
\begin{align}
    &\,
    \langle \rho(h_1)\rho(h_2)\rangle_0
    \\
    =&
    \int_{-\i\infty}^{+\i\infty}\frac{\d\beta_1}{2\pi\i}\int_{-\i\infty}^{+\i\infty}\frac{\d\beta_2}{2\pi\i} 
    \,
    \e^{\beta_1\left(h_1-\frac{c}{24}\right)+\beta_2\left(h_2-\frac{c}{24}\right)} 
    \,
    Z^{\rm primary}(\beta_1,\beta_2)
    \nn
    =&\,-\frac{\mathcal{N}}{4\pi^2}
    \frac{h_1+h_2-\frac{c-1}{12}}{\left(h_1-\frac{c-1}{24}\right)^{1/2}\left(h_1-\frac{c-1}{24}\right)^{1/2}(h_1-h_2)^2}\,\Theta\!\left(h_1+h_2-\frac{c-1}{12}-|h_1-h_2|\right)\nonumber
    \nn
    =&\,-\frac{\mathcal{N}}{4\pi^2}
    \frac{E_1+E_2}{E_1^{1/2}E_2^{1/2}(E_1-E_2)^2}\,\Theta\!\left(E_1+E_2-|E_1-E_2|\right),\nonumber
\end{align}
where we have defined $E_i\equiv h_i-(c-1)/24$, and $\Theta$ is the Heaviside step function setting $E_i>0$. When performing the integrals, it is useful to do a change of variables $\beta_{\pm}=\beta_1\pm\beta_2$. The step function comes out automatically from the integration. The result is exactly the universal cylinder amplitude for (one-cut) random matrix models \cite{Ambjorn:1990ji,Brezin:1993qg,Saad:2019lba} with GUE symmetry. Notice that the $g$-function does not appear in the expression, as desired for the universal cylinder amplitude. This is because the Euler characteristic of the EOW branes is zero. 

We have implicitly assumed that $a$ and $b$ are different boundary conditions, so there is only one way to connect the two asymptotic annuli. However, in the special case $a=b$, i.e., with identical boundary conditions, there are two ways to connect the annuli. To be more precise, the BCFT annulus partition has a CRT symmetry which exchanges the two circular borders. Similar to the closed case \cite{Yan:2023rjh}, this means that we can cut the wormhole in the middle and glue them back together after the $\mathbb{Z}_2$ mapping group class action swapping the two EOW branes:
\begin{align}\label{eq:mapZ2}
    \vcenter{\hbox{\includegraphics[height=3cm]{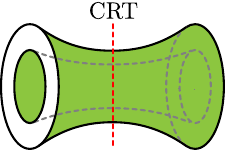}}}~.
\end{align}
This gives a factor of two for the case $a=b$ relative to when $a\ne b$, making it a GOE ensemble \cite{Bohigas:1983er}. This is in fact predicted by the matrix potential, which is a sum over $a$ and $b$, so the diagonal terms have a factor of a half relative to off-diagonal terms, which becomes a factor of two after taking the inverse \cite{Jafferis:2025yxt}.

Before moving on, we shall emphasize a point regarding the mapping class group: when considering the mapping class group, we should fix the EOW boundaries \emph{set-wise}, as they are dynamical objects in the theory. This is in contrast to asymptotic boundaries, which we fix \emph{point-wise}. In \eqref{eq:spatial}, the mapping class group of the annulus at $y=0$ is $\mathbb{Z}$ because we fix the asymptotic boundaries point-wise. When considering the annulus at $\rho=0$ as in \eqref{eq:mapZ2}, the mapping class group of the annulus is trivial (when $a\ne b$) rather than $\mathbb{Z}$ because the EOW circles are fixed set-wise.

\section{Other amplitudes}\label{sec:higher}

In a matrix model, the disk and cylinder amplitudes recursively determine all other amplitudes. Since we have already matched the disk and cylinder amplitudes with VMS, the remaining amplitudes need to be checked to ensure they are indeed described by VMS. In this section, we show that the partition function of 3d gravity on $\Sigma_{g,n}\times I$ with $\chi\equiv 2-2g-n<0$ does reproduce the VMS amplitudes on $\Sigma_{g,n}$.

In earlier sections, we were working with the moduli basis for the $n$ boundaries of $\Sigma_{g,n}\times I$, so that each of them was an asymptotic annular boundary. In this section, it is convenient to work with the momentum basis. Each asymptotic annulus then shrinks to a single line, extending between the two EOW branes. These are Wilson lines in (two copies of) Virasoro TQFT. As there are only scalar operators in VMS, we only need to consider those with $P=\bar{P}$ for the two copies.\footnote{Another reason to restrict to scalar Wilson lines in the current setup is that spinning Wilson lines intersecting EOW branes would result in ``spinning punctures" at the branes, causing an inconsistency with the BCFT ensemble interpretation of 3d gravity. For example, it would give a non-zero one-point function on the disk for a spinning operators, which should be zero in BCFT \cite{Wang:2025bcx}.}

The partition function of 3d gravity on Maldacena-Maoz (MM) wormholes  \cite{Jafferis:2025yxt}, which are topologically $M=\Sigma_{g,n}\times I$, with asymptotic boundary conditions at the ends of the interval, is equal to the Liouville CFT partition function on $\Sigma_{g,n}$ \cite{Chandra:2022bqq,Collier:2023fwi}. It is sometimes convenient to think of the Liouville partition function as a state on $\Sigma_{g,n}$ with moduli $(\mathbf{m}_1,{\mathbf{m}_2})$ prepared by two copies of Virasoro TQFT on the $\mathbb{Z}_2$ quotient of the MM wormhole \cite{Jafferis:2025yxt}:
\begin{align}\label{eq:MM}
    \vcenter{\hbox{\includegraphics[width=4cm]{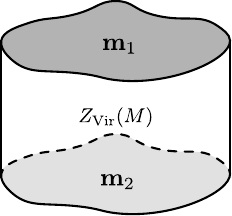}}}\quad =\quad  \vcenter{\hbox{\includegraphics[width=5cm]{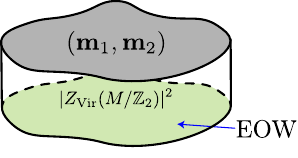}}}~.
\end{align}
It should be emphasized that the EOW brane here is not a state boundary but rather a defect in spacetime, so the $\mathbb{Z}_2$ quotient of $M$ only has one ``physical" boundary where the state lives. We will refer to $M/\mathbb{Z}_2$ as the half-MM wormhole. 

From the picture above, it is intuitively clear that what we want to compute can be obtained by taking the inner product between two such states prepared by two half-MM wormholes (using the notation for a discrete basis):
\begin{align}\label{eq:SigmaI}
    \vcenter{\hbox{\includegraphics[width=3cm]{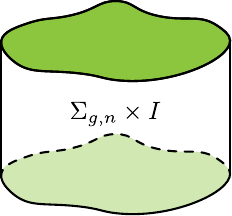}}}
    \quad=~~\sum_{i,j} G^{-1}_{ji} 
    \vcenter{\hbox{\includegraphics[width=7cm]{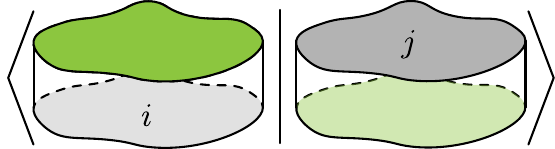}}}~,
\end{align}
where $i$ and $j$ label a complete but non-orthogonal basis and $G^{-1}$ is the inverse of the Gram matrix of inner products. 

The partition function on $\Sigma_{g,n}\times I$ is therefore given by 
\begin{align}
&\,\mathcal{Z}_{\Sigma_{g,n}\times I}(\vec{P}^{\rm ext})
\nn
=\,&(g_ag_b)^{\chi}\int_{(\mathcal{T}_{g,n}\times\mathcal{T}_{g,n})/\text{Map}(\Sigma_{g,n})}
\d^2\mathbf{m}_1
     \d^2\mathbf{m}_2\,
     Z_{\rm gh} 
     Z_{\rm tL}(\mathbf{m}_1,\bar{\mathbf{m}}_1)
     Z_{\rm gh}
     Z_{\rm tL}(\mathbf{m}_2,\bar{\mathbf{m}}_2)
\nn
&\times
     Z_{\rm Liouville}(\mathbf{m}_1,\mathbf{m}_2)
     Z_{\rm Liouville}(\bar{\mathbf{m}}_1,\bar{\mathbf{m}}_2),
\end{align}
where $Z_{\rm tL}$ is the timelike Liouville partition function, $Z_{\rm gh}$ is the ghost partition function, whose arguments have been suppressed (it always has the same moduli dependence as the $Z_{\rm tL}$ factor that comes immediately after it), $\vec{P}^{\rm ext}$ is an $n$-component real vector labeling the weights of the (scalar) Wilson lines at the boundaries (punctures) of $\Sigma_{g,n}$, $\mathcal{T}_{g,n}$ is the Teichm{\"u}ller space of $\Sigma_{g,n}$, $\text{Map}(\Sigma_{g,n})$ is the mapping class group of $\Sigma_{g,n}$, and $\chi$ is the Euler characteristic of $\Sigma_{g,n}$. The moduli parameter $\mathbf{m}$ is a $(3g-3+n)$-component complex vector, so $\d^2\mathbf{m}$ is a shorthand for $\d^{3g-3+n}\mathbf{m}\,\d^{3g-3+n}\bar{\mathbf{m}}$. Here and in the next few equations, we suppress the $\vec{P}^{\rm ext}$ dependence in places to reduce clutter.

It is important to note that this is a gravitational inner product, rather than a TQFT inner product. The difference is that the mapping class group is gauged in gravity, whose effect we have already seen in the previous section. Consequently, the integration is over the gravitational phase space  $(\mathcal{T}_{g,n}\times\mathcal{T}_{g,n})/\text{Map}(\Sigma_{g,n})$ \cite{Kim:2015qoa,Maloney:2015ina,Eberhardt:2022wlc}. (In holomorphic quantization, the inner product is given by an integral over the phase space.) 

It is useful to rewrite the phase space $(\mathcal{T}_{g,n}\times\mathcal{T}_{g,n})/\text{Map}(\Sigma_{g,n})$ as a (non-trivial) $\mathcal{T}_{g,n}$ bundle over $\mathcal{M}_{g,n}$, where we use ${\mathbf{m}}_2$ for coordinates on the fiber and ${\mathbf{m}}_1$ for coordinates on the base. At each point in the base, we then perform the integration along the fiber: 
\begin{align}
&\int_{\mathcal{T}_{g,n}} \d^2 {\mathbf{m}}_2
\,
Z_{\rm Liouville}(\mathbf{m}_1, \mathbf{m}_2) 
Z_{\rm gh}Z_{\rm t L}(\mathbf{m}_2, \bar{\mathbf{m}}_2) 
Z_{\rm Liouville}(\bar{\mathbf{m}}_1,\bar{\mathbf{m}}_2) 
\nn
=&\int_{\mathcal{T}_{g,n}} \d^2 \mathbf{m}_2
\,Z_{\rm gh}Z_{\rm t L}(\mathbf{m}_2, \bar{\mathbf{m}}_2)
\nn
&
 \times\int \d\mathbf{P}\rho_{g, n}^{\mathcal{C}}(\mathbf{P})
\mathcal{F}^\mathcal{C}_{g,n}(\mathbf{P}| \mathbf{m}_1)
\mathcal{F}^\mathcal{C}_{g,n}(\mathbf{P}| \mathbf{m}_2)
\int\d\mathbf{P}'\rho_{g, n}^{\mathcal{C}}(\mathbf{P}')
\mathcal{F}^\mathcal{C}_{g,n}(\mathbf{P}'|\bar{\mathbf{m}}_1)
\mathcal{F}^\mathcal{C}_{g,n}(\mathbf{P}'|\bar{\mathbf{m}}_2)
\nn
=
&
\int \d\mathbf{P}\rho_{g, n}^{\mathcal{C}}(\mathbf{P})
\mathcal{F}^\mathcal{C}_{g,n}(\mathbf{P}| \mathbf{m}_1)
\int\d\mathbf{P}'\rho_{g, n}^{\mathcal{C}}(\mathbf{P}')
\mathcal{F}^\mathcal{C}_{g,n}(\mathbf{P}'|\bar{\mathbf{m}}_1)
\frac{\delta{(\mathbf{P}-\mathbf{P}')}}{\rho_{g, n}^{\mathcal{C}}(\mathbf{P})}
\nn
=
&
\int \d\mathbf{P}\rho_{g, n}^{\mathcal{C}}(\mathbf{P})
\mathcal{F}^{\mathcal{C}}_{g,n}(\mathbf{P}| \mathbf{m}_1)
\mathcal{F}^{\mathcal{C}}_{g,n}(\mathbf{P}|\bar{\mathbf{m}}_1)
=
Z_{\rm Liouville}({\mathbf{m}}_1,\bar{\mathbf{m}}_1),
\end{align}
where in going from the second to the third line we used the Verlinde inner product \cite{Verlinde:1989ua,Collier:2023fwi} between Liouville blocks
\begin{align}
    \int_{\mathcal{T}_{g,n}}
    \mathcal{F}^\mathcal{C}_{g,n}(\vec{P}^{\rm ext},\mathbf{P}|\mathbf{m})
    \mathcal{F}^\mathcal{C}_{g,n}(\vec{P}^{\rm ext},\mathbf{P}'|\bar{\mathbf{m}})
    =
    \frac{\delta{(\mathbf{P}-\mathbf{P}')}}{\rho_{g, n}^{\mathcal{C}}(\mathbf{P})},
\end{align}
and used the compact notation \cite{Collier:2023fwi}
\begin{align}
\rho_{g, n}^{\mathcal{C}}(\mathbf{P})\equiv \prod_{\text {cuffs } a} \rho_0(P_a) \prod_{\substack{\text { pairs of pants } \\(i, j, k)}} C_0(P_i, P_j, P_k),
\end{align}
$\mathcal{C}$ being a choice of channel decomposition, $\mathbf{P}$ being the internal momenta, $\rho_0$ being the Cardy density, and $C_0$ being the DOZZ formula \cite{Dorn:1994xn,Zamolodchikov:1995aa} up to some normalization \cite{Collier:2019weq}.

Using this relation, the gravitational inner product simplifies, and we conclude that
\begin{align}
\mathcal{Z}_{\Sigma_{g,n}\times I}(\vec{P}^{\rm ext})
&=(g_ag_b)^{\chi}\int_{\mathcal{M}_{g,n}}\d^2 \mathbf{m}_1\,
    Z_{\rm gh}(\mathbf{m}_1,\bar{\mathbf{m}}_1) Z_{\rm tL}(\mathbf{m}_1,\bar{\mathbf{m}}_1)Z_{\rm Liouville}(\mathbf{m}_1,\bar{\mathbf{m}}_1)
\nn
&=\mathsf{V}_{g,n}(\vec{P}^{\rm ext})
\end{align}
is exactly the VMS amplitude on $\Sigma_{g,n}$ with external momenta $\vec{P}^{\rm ext}$, also known as the quantum volume. Here, by our convention, we have kept the $\e^{S_0}$ dependence in the VMS expressions. From this, we identify $\e^{S_0}=g_ag_b$.

Finally, to establish the exact correspondence, we need to show that the matrix spectral correlators computed from $\mathcal{Z}_{\Sigma_{g,n}\times I}(\vec{P}^{\rm ext})$ using \eqref{eq:dualVMSitoZ} agree with those computed from $\mathsf{V}_{g,n}(\vec{P}^{\rm ext})$, using the duality between VMS and the matrix model. This requires an analysis of the trumpet partition function. Adding the trumpet amounts to
\begin{align}
    \mathcal{Z}_{\Sigma_{g,n}\times I}(\i t_1,\dots,\i t_n)
    &=\int _0^\infty\prod_i\left[ \d(P_i^2)\,\mathcal{Z}^{\rm trumpet}_{\Sigma_{g,n}\times I}(\i t_i,P_i)
    \right]
    \mathcal{Z}_{\Sigma_{g,n}\times I}(P_1,\dots,P_n)
\end{align}
where $\vec{P}^{\rm ext}=(P_1,\dots,P_n)$, and 
\begin{align}
    \mathcal{Z}^{\rm trumpet}_{\Sigma_{g,n}\times I}(\i t_i,P_i)
    &=
    \tilde{\mathcal{Z}}^{\rm trumpet}_{\Sigma_{g,n}\times I}(\i/ t_i,P_i)
    \nn
    &=\frac{1}{\eta(\i/t_i)}
    \e^{-{2\pi P^2}/{t_i} }
    \nn
    &=\frac{1}{\sqrt{t_i}\,\eta(\i t_i)}
    \e^{-{2\pi P^2}/{t_i} }.
\end{align}
Here, $\tilde{\mathcal{Z}}^{\rm trumpet}_{\Sigma_{g,n}\times I}$ is the trumpet partition function when the boundary is in the closed-string channel, and we have used \eqref{eq:open-closed} to relate it to ${\mathcal{Z}}^{\rm trumpet}_{\Sigma_{g,n}\times I}$.

Stripping off the descendant contributions using \eqref{eq:stripeq}, we get
\begin{align}\label{eq:Zprim_ZVMS}
    \mathcal{Z}^{\rm primary}_{\Sigma_{g,n}\times I}(\i t_1,\dots,\i t_n)
    &=\int _0^\infty \prod_i\left[\d(P^{\prime2}_i)
    \frac{1}{\sqrt{t_i}\,\e^{-\pi t_i/12}}
    \e^{-{2\pi P_i'^2}/{t_i} }
    \right]
    \mathsf{V}_{g,n}({P}^{\prime}_1,\dots,P'_n)
    \nn
    &=\e^{(\beta_1+\cdots+\beta_n)/24}
   \,\mathsf{Z}_{g, n}(\beta_1, \ldots, \beta_n),
\end{align}
where $\mathsf{Z}_{g, n}\left(\beta_1, \ldots, \beta_n\right)$ is the VMS partition function on $\Sigma_{g,n}$ with (circular) asymptotic boundaries with inverse temperatures $\beta_i$, with the $\e^{S_0}$ dependence kept \cite{Collier:2023cyw}. 

The spectral correlators defined via \eqref{eq:dualVMSitoZ} then become
\begin{align}
    \langle \rho(h_1)\cdots \rho(h_n)\rangle_g
    &=
    \prod_{i=1}^{n}
    \int_{\gamma-\i\infty}^{\gamma+\i\infty}
    \left[
    \frac{\d\beta_i}{2\pi\i} \e^{\beta_i H_i}
    \right]
    \e^{(\beta_1+\cdots+\beta_n)/24}
   \,\mathsf{Z}_{g, n}(\beta_1, \ldots, \beta_n)
   \nn
   &=    
    \prod_{i=1}^{n} \int_{\gamma-\i\infty}^{\gamma+\i\infty}
    \left[
    \frac{\d\beta_i}{2\pi\i} \e^{\beta_i \left(h_i-\frac{c-1}{24}\right)}
    \right]
   \mathsf{Z}_{g, n}(\beta_1, \ldots, \beta_n)
   \nn
   &=\langle \varrho(E_1)\cdots \varrho(E_n)\rangle_g^{\rm VMS}|_{E_i\to h_i-(c-1)/24}.
\end{align}
This concludes the derivation. It is interesting to note that the partition functions themselves do not exactly agree, but differ by a simple factor shown in \eqref{eq:Zprim_ZVMS}. This simple factor exactly combines with the Casimir energy of the BCFT to turn the Hamiltonian $H$ into the Liouville momentum $E=P^2$, the conjugate variable to $\beta$ in VMS.

\section{Discussion}\label{sec:disc}

In conclusion, we have established a relation between the VMS matrix model and 3d gravity on $\Sigma_{g,n}\times I$ with EOW branes. We now provide several comments and discuss some open questions.

Throughout this work, we have frequently used the doubling trick, relating 3d gravity with EOW branes on $\Sigma_{g,n}\times I$ to a chiral copy of $\SL$ Chern-Simons theory on $\Sigma_{g,n}\times S^1$. We should emphasize that this is only a trick for several reasons. One of them is that the EOW branes generally have tension, and the partition function depends on the tension through the topology of the brane \cite{Geng22,Wang:2025bcx}. On the brane, this is a 2d topological theory equivalent to that of \cite{Marolf:2020xie}. This ensures the existence of a topological expansion parameter from the matrix model perspective. Chiral 3d gravity does not have this topological dependence. Moreover, in the metric formalism, 3d gravity with EOW branes makes sense locally, as the metric is physical everywhere. Another reason is the interpretation from the 2d CFT perspective. In our case, the BCFT annulus partition function plays a central role. In particular, the Hamiltonian provides a physical motivation for performing the inverse Laplace transforms.

For the cylinder amplitude, we performed a \emph{bona fide} computation for the partition function of the annulus wormhole, $S^1\times I\times I$, where one of the intervals ends at asymptotic boundaries and the other ends at EOW branes. The computation is analogous to that of the torus wormhole \cite{Cotler:2020ugk}, but we did not employ any formal step to obtain a measure by examining the action. It would be interesting to do the same for the torus wormhole. We have also explicitly demonstrated how the mapping class group renders the answer finite. 

For amplitudes other than the disk and the cylinder, we should contrast the derivation with that of \cite{Collier:2023cyw}, which established a connection between VMS and chiral 3d gravity. First of all, we are computing an inner product rather than a formal trace over a Hilbert space. The inner product involves an integration over the well-defined phase space of gravity, whereas the expression using the formal trace is technically infinity divided by infinity. Furthermore, the mapping class group of $\Sigma_{g,n}\times I$ is the same as that of $\Sigma_{g,n}$, unlike $\Sigma_{g,n}\times S^1$, so the argument of \cite{Collier:2023cyw} requires some additional assumptions about the gauge symmetry of chiral 3d gravity. Finally, the correspondence between gravity with EOW branes and the matrix model is exact for the spectral correlators, while the partition functions in the moduli basis differ by a simple factor. The factor has a good physical explanation from the BCFT perspective.

We have restricted our attention to a large class of off-shell manifolds with EOW branes, namely $\Sigma_{g,n}\times I$, and shown that they are related to a random matrix model. Other off-shell manifolds can be obtained by starting with on-shell manifolds and gluing these manifolds to them \cite{Jafferis:2024jkb}. In the purely open tensor-matrix model of \cite{Jafferis:2025yxt}, on-shell manifolds with EOW branes and boundary Wilson loops are generated by the tensor part of the model's potential. We then remove the neighborhood of each of the Wilson loops, leaving an annular state-cut boundary. At the leading order, the thickened disk \eqref{eq:BTZcut} is glued to each of such annuli, a procedure also known as the annular surgery. For any set of $n$ annuli, we glue $\Sigma_{g,n}\times I$ to it for every $g$ to obtain higher-order corrections. This makes predictions for the partition functions of manifolds with more complicated off-shell topology, and it would be interesting to check them.

A natural next step is to do the analogous analysis for the purely closed tensor-matrix model \cite{Belin:2023efa,Jafferis:2024jkb} or extend this to the full open-closed tensor-matrix model \cite{Jafferis:2025yxt}. The immediate challenge is to generalize the main conclusions of this paper to the closed matrix model, describing 3d gravity on $\Sigma_{g,n}\times S^1$. Related approaches to this problem have been pursued, for example in \cite{DiUbaldo:2023qli,Boruch:2025ilr,deBoer:2025rct}. It would be interesting to explore the connection.

We should also point out that BCFT plays another role in the story. In this work, we have used the term ``BCFT" to exclusively refer to the asymptotic boundaries of 3d gravity, but in the worldsheet description of VMS, BCFT is used to describe for example the trumpet and instantons. We have not needed to discuss them explicitly, partly because we have already taken as an input the duality between the worldsheet VMS and the matrix model established in \cite{Collier:2023cyw}. One should not confuse the two separate roles played by BCFT.

Finally, it is noteworthy that when the asymptotic boundary is a single annulus, there is no $\mathrm{SL}(2,\mathbb{Z})$ modular sum like the torus case. An annulus that is homotopic to the asymptotic boundary (say a constant-radius slice) has trivial mapping class group (since it has EOW boundaries, as explained at the end of Section~\ref{sec:cylinder}). The analog of the Maloney-Witten sum \cite{Maloney:2007ud,Keller:2014xba} is therefore just the Cardy density times the $g$-functions (plus the vacuum state when the boundary conditions are the same). This spectral density is manifestly positive.

\section*{Acknowledgements}

It is a pleasure to thank Jeevan Chandra, Scott Collier, Charlie Cummings, Gabriele Di Ubaldo, Tom Hartman, Luca Iliesiu, Alex Maloney, Julian Sonner, Joaquin Turiaci, Zixia Wei, Cynthia Yan, and Mengyang Zhang for helpful discussions. 
DLJ, LR and DS acknowledge support by the Simons Investigator in Physics Award MP-SIP-0001737, U.S. Department of Energy grant DE-SC0007870, and Harvard University. DW acknowledges support by NSF grant PHY-2207659 and the Simons Collaboration on Celestial Holography. 

\appendix

\bibliographystyle{JHEP}
\bibliography{library}

\end{document}